\documentstyle[aps,prl,epsf,twocolumn]{revtex}
\begin{document}
\draft
\title{How to identify a Strange Star}
\author{Jes Madsen}
\address{Institute of Physics and Astronomy, University of Aarhus, 
DK-8000 \AA rhus C, Denmark}
\date{June 2, 1998}
\maketitle

\begin{abstract}
Contrary to young neutron stars, young strange stars are not subject to the
$r$-mode instability which slows rapidly rotating, hot neutron stars to 
rotation periods near 10 ms via gravitational wave emission. Young
millisecond pulsars are therefore likely to be strange stars rather
than neutron stars, or at least to contain significant quantities of
quark matter in the interior.
\end{abstract}


It has recently been shown, that the emission of gravitational radiation
due to $r$-mode instabilities in hot, young neutron stars severely
limits the rotation period of these stars 
\cite{and98,frimor98,linowe98,andkok98,kokste98,oweal98}. 
Within the first year, where
the star cools to $10^9$K, the star will spin down to a few per cent of
the Kepler-limit \cite{kepler} 
(the ultimate rotation frequency where the stellar equator
``falls off''), corresponding to a rotation period of 10--20
milliseconds. This scenario has several interesting ramifications
\cite{and98,frimor98,linowe98,andkok98,kokste98,oweal98}. First
of all, the ``slow'' rotation predicted for young pulsars seems to agree
with the few observed periods of pulsars known to be young, such as
Crab (the rapid millisecond pulsars with periods as small as 1.56 ms are
expected to be old, cold pulsars spun-up by angular momentum accretion in
binary systems, and are therefore not subject to the $r$-mode
instability). Second, the spin-down involves emission of up to
$10^{52}$ergs of gravitational waves, making the
radiation potentially observable. And third, the instability rules out
a scenario where millisecond pulsars form by accretion-induced collapse
of a white dwarf, because the temperature during such a process will be
high enough for the $r$-mode instability to lead to spin-down.

All rotating relativistic stars are generically unstable against
the $r$(otational)-mode instability \cite{and98,frimor98}. 
To first order in the rotational
frequency $\Omega$ of the star, the frequency observed by a distant
inertial observer is $\omega_i = \omega_r -m\Omega$, where the frequency
in the rotating frame is $\omega_r=2m\Omega/l(l+1)$ for a mode with
spherical harmonic indices $(l,m)$. Thus modes which are
counterrotating in the comoving frame will appear corotating to a
distant observer. Therefore, $r$-modes are
generically unstable to the emission of gravitational waves; the
emission removes positive angular momentum from a mode with
negative angular momentum in the corotating frame, thereby making the
angular momentum and energy increasingly negative. In contrast to other known
modes, the $r$-mode instability does not require a large rotation rate,
but sets in even as the rotation frequency goes to zero, at least in a
non-viscous system (for a review of instabilities in rotating
relativistic stars the reader is referred to \cite{ste98}).

Important for the $r$-mode instability scenario is the effect of
internal fluid dissipation in the star that tends to suppress the
instabilities. Lindblom, Owen, and Morsink \cite{linowe98}, 
and Andersson, Kokkotas, and Schutz \cite{andkok98}
have shown that for typical neutron star equations of state, the
shear viscosity limits the $r$-mode instability at temperatures below $10^9$K,
whereas bulk viscosity is the limiting factor above this temperature.
And they find that the rotation rate for
standard assumptions about neutron star cooling is
reduced to 7--8\% of the Keplerian limit.

Whereas the shear viscosity of quark matter \cite{heipet93} 
is roughly comparable to that of
neutron star matter for the parameter range of interest here,
the bulk viscosity of quark matter is larger by many orders
of magnitude \cite{mad92}. 
Therefore, one should expect the $r$-mode instability to
be significantly suppressed if the stars are made up of $u$-, $d$-, and
$s$-quarks (so-called strange stars \cite{wit84}), 
rather than ordinary neutron star
matter. A similar result would apply if quark matter is metastable
rather than absolutely stable, so that a neutron star contains quark
phase in its central regions. 

In the following I will show that this is
indeed the case. The $r$-mode instability does not play any role in a
young strange star. Therefore, finding a young pulsar with a rotation
period below 5--10 milliseconds would be a strong indication of the
existence of strange stars, or at least a significant quark content in
neutron stars, and therefore of the stability or metastability of
bulk strange quark matter. In addition it would re-introduce the
possibility of forming some millisecond pulsars by accretion-induced collapse
of white dwarf stars. Unfortunately it would also remove $r$-mode
instability braking of young neutron stars as a strong source of
gravitational radiation.

The shear viscosity of strange quark matter due to quark scattering was
calculated in \cite{heipet93}. The results for 
$T\ll \mu $, where $T$ is the temperature, and $\mu\approx 300 {\rm
MeV}$ the quark chemical potential, can be written as 
\begin{equation}
\eta \approx 1.7\times 10^{18}\left( {\frac{{0.1}}{{\alpha _S}}}
\right)^{5/3} \rho_{15}^{14/9}T_9^{-5/3}\, {\rm gcm}
^{-1}{\rm s}^{-1},
\end{equation}
where $\alpha_S$ is the fine structure constant of the strong
interactions, $T_9\equiv T/10^9 {\rm K}$, and $\rho_{15}\equiv \rho/10^{15}
{\rm g cm}^{-3}$. 

The bulk viscosity of strange quark matter \cite{mad92} 
depends mainly on the rate of the non-leptonic weak interaction \cite{rate}
\begin{equation}
u+d\leftrightarrow s+u.  \label{conver}
\end{equation}
Because the strange quark is much more massive than up and down quarks,
this reaction changes the concentrations of down and strange quarks in response
to the density changes involved in vibration or rotational instabilities,
thereby causing dissipation. This dissipation is most efficient if the rate
of reaction (\ref{conver}) is comparable to the frequency of the density
change. If the weak rate is very small, the quark concentrations keep their
original values in spite of a periodic density fluctuation, whereas a very
high weak rate means that the matter immediately adjusts to follow the true
equilibrium values reversibly. But in the intermediate range dissipation due
to $PdV$-work is important.

The importance of dissipation due to Eq.~(\ref{conver}) was first stressed
by Wang and Lu \cite{wanlu84} in the case of neutron stars with quark cores.
These authors made a numerical study of the evolution of the vibrational
energy of a neutron star with an $0.2M_{\odot }$ quark core, governed by the
energy dissipation due to Eq.~(\ref{conver}). Sawyer \cite{saw89} expressed
the damping in terms of the bulk viscosity, a function of temperature and
oscillation frequency, which he tabulated for a range of densities and
strange quark masses. This tabulation has later been used in studies of
quark star vibration \cite{cutlin90}, and of the gravitational radiation
reaction instability in rapidly rotating quark stars \cite
{colmil92}. The latter study concluded, that the bulk viscosity was large
enough to be important for temperatures exceeding $10^8$K, but that it
should be a few orders of magnitude larger to generally dominate the
stability properties.

However, as pointed out in Ref.\ \cite{mad92}, the bulk
viscosities in \cite{saw89} depend on the assumption, that the rate of
Eq.~(\ref{conver}) can be expanded to first order in $\delta \mu =\mu _s-\mu
_d$, where $\mu _i\approx 300$MeV are the quark chemical potentials. This
assumption is not correct at low temperatures ($2\pi T\ll \delta \mu $),
where the dominating term in the rate is proportional to $\delta \mu ^3$.
Furthermore, the rate in \cite{saw89}
is too small by an overall factor of 3, and another
discrepancy of 2--3 orders of magnitude, perhaps due to unit conversions,
appears as well. Taken together, these effects led to an upward correction
of the bulk viscosity by several orders of magnitude, and thereby increases
the importance for the astrophysical applications. The non-linearity of the
rate also means, that the bulk viscosity is no longer independent of the
amplitude of the density variations. The resulting bulk viscosity is (in
cgs-units, with strange quark mass $m_s$, $T$, 
and $\mu _d\approx 359{\rm MeV}\rho_{15}^{1/3}$
in MeV, and the frequency of the perturbation $\omega $ in s$^{-1}$)
\begin{eqnarray}
\zeta \approx &&1.10\times 10^{29}m_s^4\omega ^{-2} \rho_{15}\nonumber \\
&& \times \left[ {\frac 34}\left( {\frac{m_s^2}{{3\mu _d}}}{\frac{{%
\Delta v}}{{v_0}}}\right) ^2+4\pi ^2T^2\right] {\rm g\,cm}^{-1}{\rm s}^{-1}.
\label{bulkvis}
\end{eqnarray}
For typical values ($m_s=100$ MeV, $\mu_d=300$MeV, $\omega = \omega_r=
2m\Omega/l(l+1) =2\times 10^3$ s$ ^{-1}$) 
this is $\zeta \approx 1.6\times 10^{30}\left[ 93(\Delta
v/v_0)^2+0.29T_9^2\right] {\rm g\,cm}^{-1}{\rm s}^{-1}$, where $\Delta v/v_0$ is
the perturbation amplitude.

Kokkotas and Stergioulas \cite{kokste98} give an analytic description of
the $r$-mode instability in uniform density neutron stars, finding
results in good agreement with more sophisticated numerical calculations. The
approximation of constant density is even better for a strange star,
except very close to the gravitational instability limit, so for the
estimates in the present investigation it is safe to use their
approximate formulae for damping times etc. In particular,
the damping time due to shear viscosity is given by \cite{kokste98}
\begin{equation}
\tau_{\rm sv} = {{3}\over{4\pi(l-1)(2l+3)}}{{M}\over{\eta R}},
\end{equation}
and the corresponding timescale for damping due to bulk viscosity
\begin{equation}
\tau_{\rm bv} = {{3(2l+5)}\over{2\pi(l+1)^3}}{{\Gamma^4 M}\over{\zeta R}},
\end{equation}
with $\Gamma\approx 5$ to simulate an almost uniform density.

The growth time due to emission of gravitational waves 
($\tau_{\rm gw}=-2E/(dE/dt)|_{\rm gw}$) is to lowest order in the angular
velocity, $\Omega$, given by
\begin{equation}
\tau_{\rm gw}=-{{c^{2l+3}}\over{24G}}{{[(2l+3)!!]^2}\over{(2l+3)(l-1)^{2l}}}
\left({{l+1}\over{l+2}}\right)^{2l+2}{{\Omega^{-2l-2}}\over{MR^{2l}}} .
\end{equation}
If the timescale for gravitational wave emission is short compared to
the damping timescales $\tau_{\rm sv}$ and $\tau_{\rm bv}$, the star
will spin down. 

To find the critical angular velocity for a given stellar model as a function
of temperature one solves the equation
\begin{equation}
{{1}\over{\tau_{\rm gw}}}+{{1}\over{\tau_{\rm sv}}}+{{1}\over{\tau_{\rm
bv}}} =0 .
\label{limit}
\end{equation}
In the case of hot neutron stars, the $l=2$, $m=2$ $r$-mode instability
was found to be decisive in the sense that it corresponds to the longest
viscous timescales. 

Introducing $R_{10}\equiv R/10 {\rm km}$, 
$M_{1.5}\equiv M/1.5 M_\odot$, where $M_\odot$ is the solar mass, and
$\Omega_3\equiv \Omega/10^3 {\rm s}^{-1}$, one gets 
\begin{equation}
\tau_{\rm gw} =-1.29\times 10^6 \, {\rm s}\, 
\Omega_3^{-6} M_{1.5}^{-1} R_{10}^{-4}.
\end{equation}
For strange quark matter the viscous timescales are
\begin{equation}
\tau_{\rm sv}({\rm SQM}) = 1.01\times 10^8\, {\rm s}\, (\alpha_S/0.1)^{5/3}
M_{1.5}^{-5/9} R_{10}^{11/3} T_9^{5/3},
\end{equation}
\begin{equation}
\tau_{\rm bv}({\rm SQM}) = 5.75\times 10^{-2}\, {\rm s}\, 
m_{100}^{-4} \Omega_3^2 R_{10}^2 T_9^{-2} .
\end{equation}
Here $m_{100}\equiv m_s/100{\rm MeV}$, and for simplicity 
only the $T$-dependent term,
not the $\Delta v/v_0$-term in Eq.\ (\ref{bulkvis}) has been included.

For comparison the shear viscosity in ordinary neutron star matter is
dominated by electron-electron scattering for $T_9<1$. The
corresponding timescale is \cite{kokste98}
\begin{equation}
\tau_{\rm sv}({\rm NS}) = 3.4\times 10^7 \, {\rm s}\, 
M_{1.5}^{-1} R_{10}^5 T_9^2.
\end{equation}
If the proton fraction in the neutron star is below $\frac{1}{9}$, the
modified URCA-process is the dominant weak interaction, giving a bulk
viscosity damping time \cite{kokste98}
\begin{equation}
\tau_{\rm bv}({\rm NSmU}) = 4.33\times 10^9 \, {\rm s}\, \Omega_3^2
M_{1.5}^{-1} R_{10}^5 T_9^{-6} .
\end{equation}
If a larger proton fraction is available, as happens in some equations
of state, the faster direct URCA-process is active \cite{latal91}, and the bulk
viscosity timescale changes to 
\begin{equation}
\tau_{\rm bv}({\rm NSdU}) = 227 \, {\rm s}\, (qx_{01})^{-1/3} \Omega_3^2
M_{1.5}^{2/3} T_9^{-4} ,
\end{equation}
where $qx_{01}\approx 1$ (c.f.\ \cite{zdu96}).

Inserting these timescales in Eq.\ (\ref{limit}) one can solve for a
critical angular velocity, $\Omega_c$, for the onset of the $r$-mode
instability. Figure 1 shows the results as functions of temperature for
stellar models with $M_{1.5}=R_{10}=1$. $\Omega_c$ is expressed in units
of the maximum (Kepler) angular velocity, $\Omega_K\approx 0.67\sqrt{\pi
G\rho}\approx 8185 {\rm s}^{-1} M_{1.5}^{1/2}R_{10}^{-3/2}$.

For neutron stars with proton fraction below $\frac{1}{9}$ 
the transition between shear- and bulk
viscosity damping is seen to take place at $T\approx 10^9$K, whereas
direct URCA-processes reduces the transition temperature to $10^8$K. The
time required to cool the interior of a neutron star to $10^9$K by the
modified URCA mechanism is less than a year, and via the direct URCA
mechanism, the cooling timescale is roughly $t_{\rm cool}\approx 20
{\rm s} T_9^{-4}$ (see \cite{pet92} for a review of neutron star
cooling), so in any case, the neutron star will be subject to
significant $r$-mode damping of the rotation rate within a matter of days to
months.

In contrast, the transition between shear- and bulk viscosity damping in
a strange star only takes place after the interior of the star has
cooled to around $4.6\times 10^6 {\rm K} (\alpha_S/0.1)^{10/19}
m_{100}^{-18/19}$, and with a reduction in $\Omega_c$ to
between 0.26 and 0.32 of the Kepler frequency for a strange quark mass
of 100 and 200 MeV respectively, corresponding to rotation periods of 3
milliseconds or faster, rather than the much stronger reductions
experienced by a neutron star. Furthermore, the time required for a
strange star to cool to the temperatures where the $r$-mode instability
rotation damping sets in is much longer than the few months in the case
of a neutron star. A typical cooling timescale for a strange star 
is $t_{\rm cool}\approx 10^{-4} {\rm years} T_9^{-4}$, so that cooling
to $4.6\times 10^6 {\rm K}$ would take $2\times 10^5$ years (and if the
electron fraction in the star is very low, cooling may have to proceed
on a much longer timescale, the quark matter analog of the modified
URCA process; see \cite{schal97} for a recent study of strange star
cooling). 

If the initial rotation rate of the strange star is very
rapid, say half the Kepler frequency, the $r$-mode instability will set
in after a few years for low strange quark mass, but increasing the 
rotation period
even above 2 ms will take more than $10^3$ years (and to reach the
ultimate limit of 3 ms takes at least a hundred times longer).

These results were derived keeping only the
temperature dependent term in the bulk viscosity of strange quark
matter, Eq.\ (\ref{bulkvis}) \cite{dailu96}. The actual viscosity
is even higher due to the presence of the nonlinear term depending on the
amplitude of the density perturbation. Thus the bulk viscosity will
dominate at even lower temperatures than found here, but a detailed
numerical study of the eigenmodes in a stellar model is required to
include the nonlinear term properly.

It therefore seems safe to conclude, that strange stars differ
significantly from ordinary neutron stars in terms of their (lack of)
sensitivity to the spin-down due to the $r$-mode instability. A young,
rapidly rotating pulsar with period below 5--10 milliseconds could
therefore very well be a strange star, if strange quark matter is
absolutely stable, or a neutron star with a significant core of strange
quark matter, if strange matter is only metastable (a so-called hybrid
star) \cite{loop}. At present no such
young millisecond pulsars (i.e.\ young pulsars with periods below 10
milliseconds) are known, which may either indicate, that pulsars are
neutron stars rather than quark stars, or perhaps that the
stars are all formed with low angular momenta for some other reason. 
Old neutron stars
being spun up by angular momentum transfer in binaries are not subject
to the $r$-mode instability, so the presently known millisecond pulsars
could be either neutron or strange stars, whereas an old submillisecond
pulsar because of the extremely high bulk viscosity of strange quark
matter might be a strange star but probably not a neutron star
\cite{saw89,colmil92,mad92}. 

More detailed calculations will be necessary to set the exact limits for
the rotation frequency of neutron stars, hybrid stars and quark stars as
a function of age and cooling history, but the discovery of even a
single young pulsar in the millisecond regime would be most exciting for
the strange quark matter hypothesis.

\acknowledgments
This work was supported in part by the Theoretical Astrophysics Center
under the Danish National Research Foundation.

\begin{figure}
\epsfxsize=7.4truecm\epsfbox{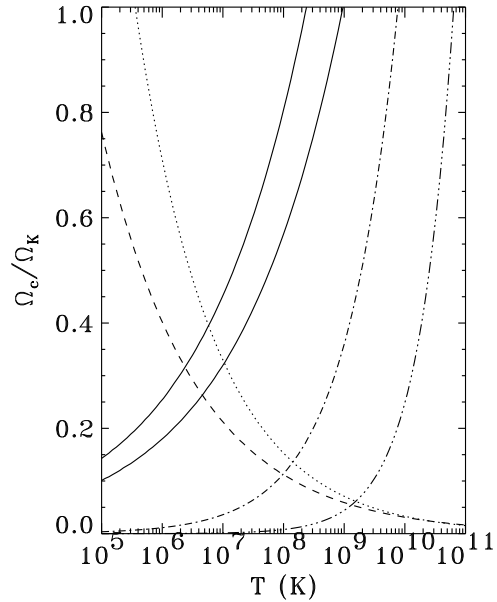}
\caption{Critical angular frequencies in units of the Kepler frequency
as functions of temperature derived from Eq.\ (\protect\ref{limit}).
Dotted curve corresponds to the shear viscosity for neutron stars,
dashed curve to shear viscosity for strange stars. Dashed-triple-dot
curve includes bulk viscosity from modified URCA processes in neutron
stars; dash-dot curve bulk viscosity from direct URCA processes in
neutron stars, and full curves are based on bulk viscosity of strange
quark matter for $m_s=100$MeV (lower curve), and $m_s=200$MeV (upper
curve). The $r$-mode instability is active when
simultaneously above the relevant curves for shear and bulk viscous
damping.}
\end{figure}


\begin{references}
\bibitem{kepler} For a Newtonian, uniformly rotating, uniform density
star the maximal angular velocity is $\Omega_K\approx 0.67\sqrt{\pi
G\rho}\approx 8185 {\rm s}^{-1} (M/1.5M_\odot )^{1/2}(R/10{\rm km})^{-3/2}$.
\bibitem{loop} Another interesting possibility would be that some other
exotic phase in the neutron star led to a dramatic increase in the
viscosity. The direct URCA process is an example of a process that does
increase the bulk viscosity, but in a less dramatic way.
Nuclear matter with high
strangeness content or even more exotic phases might experience
something similar. However, since such
phases are typically concentrated near the center of the star, nothing
as dramatic as seen for a strange star with high viscosity quark phase
almost to the surface should be expected.
\end{references}
\end{document}